\documentclass[9pt,journal,compsoc]{IEEEtran}

\usepackage{graphicx}

\usepackage[cmex10]{amsmath}
\usepackage{amssymb}
\usepackage{amsthm}
\usepackage{diagbox}

\usepackage[binary-units=true]{siunitx}
\usepackage{textcomp}
\DeclareMathSymbol{\varOmega}{\mathalpha}{operators}{"0A}
\providecommand*{\upOmega}{\varOmega}%

\usepackage{verbatim}
\ifCLASSOPTIONcompsoc
  \usepackage[nocompress]{cite}
\else
  \usepackage{cite}
\fi
\newcommand{\udl}{\underline}

\newcommand{\eps}{\epsilon}

\newcommand{\pa}[1]{\left( #1 \right)}

\newcommand{\beq}[2]{\begin{equation}\label{#1} #2 \end{equation}}
\newcommand{\bal}[2]{{\setlength\arraycolsep{2pt}\begin{eqnarray}\label{#1} #2 \end{eqnarray}}}

\newcommand{\iid}{i.i.d. }

\begin{document}
\title{Implementation and Analysis of Stable PUFs Using Gate Oxide Breakdown}

\author{Wei-Che Wang, Yair Yona, Yizhang Wu, Suhas Diggavi and Puneet Gupta}


\IEEEtitleabstractindextext{%
\begin{abstract}
We implement and analyze highly stable PUFs using two random gate oxide breakdown mechanisms: plasma induced breakdown and voltage stressed breakdown. These gate oxide breakdown PUFs can be easily implemented in commercial silicon processes, and they are highly stable. We fabricated bit generation units for the stable PUFs on 99 testchips with 65nm CMOS bulk technology. Measurement results show that the plasma induced breakdown can generate complete stable responses. For the voltage stressed breakdown, the responses are with 0.12\% error probability at a worst case corner, which can be effectively accommodated by taking the majority vote from multiple measurements. Both PUFs show significant area reduction compared to SRAM PUF. We compare methods for evaluating the security level of PUFs such as min-entropy, mutual information and guesswork as well as inter- and intra-FHD, and the popular NIST test suite. We show that guesswork can be viewed as a generalization of min-entropy and mutual information. In addition, we analyze our testchip data and show through various statistical distance measures that the bits are independent. Finally, we propose guesswork as a new statistical measure for the level of statistical independence that also has an operational meaning in terms of security.
\end{abstract}

\begin{IEEEkeywords} Hardware Security, Gate Oxide Breakdown, Guesswork Analysis
\end{IEEEkeywords}}

\maketitle
\IEEEdisplaynontitleabstractindextext
\IEEEpeerreviewmaketitle

\IEEEraisesectionheading{\section{Introduction}\label{sec:introduction}}
\IEEEPARstart{P}{hysical} Unclonable Functions (PUFs) \cite{Blaise02} have been considered as promising security primitives for the Internet of Things (IoT) for its lightweight hardware implementation. PUFs can be exploited in a variety of applications, such as identification \cite{Krishna2011} or secret key generation. The randomness of a PUF is extracted from random uncontrollable process variations, and its behavior, or Challenge Response Pair (CRP) \cite{Roel2010}, is uniquely tied to a given device and is hard to predict or replicate. Since the first physical unclonable identification was fabricated in \cite{Lofstrom2000}, extensive efforts have been devoted into the area, and different silicon PUF implementations have been proposed, including Arbiter PUF \cite{LeeJW04}, Ring Oscillator (RO) PUF \cite{Maiti2009}, SRAM PUF \cite{Holcomb09}, and many other variations. \footnote{This work is a significant extension of \cite{Wang17}}

The instability of a parametric PUF potentially limits the practical application of a PUF. Since these PUFs are parametric, they are in nature susceptible to environmental variations, and the behavior of a PUF can be altered consistently in two different environments. To make a PUF more stable, extra overhead is required, including hardware or latency cost \cite{Majzoobi2010}. Techniques such as error correction code (ECC) or helper data come with the cost of extra hardware implementation or possible security concerns \cite{Jeroen2014}.

Recently, a stability-guaranteed Locally Enhanced Defectivity PUF (LEDPUF) proposed in \cite{Wang16} shows completely stable responses by utilizing random hard defect generated from Directed Self Assembly (DSA) process. However, it is difficult to fabricate given that DSA is not well accepted into commercial silicon manufacturing yet. In \cite{RLiu2016} a reliable RRAM PUF with actual PUF fabrication using Resistive Random Access Memory (RRAM) is presented. However, an off-chip characterization of the split current and offset for the sense amplifiers are required, and the reliability results under voltage variations are not reported, which can dramatically impact the stability. Another reliable PUF using Hot Carrier Injection (HCI) is presented in \cite{Bhargava2013}. However, post calibration steps are still needed and the randomness of the most stable responses was not reported. In \cite{Liu10}, the authors apply high voltage to induce gate oxide breakdown of transistors to extract stable randomness. However, a "afterburn" phase is performed to all broken oxides to enhance the stability, which would require additional hardware and calibration. In \cite{Tang2014}, the authors intentionally introduce oxide breakdown by violating antenna rules to generate stable random bits. However, the response time may be long due to the limited leakage current to charge the ID generation output if no breakdown occurs.

A PUF designer needs to meet a desired security level without using an excessive number of gates. The two most dominant factors that can reduce the security level of a PUF are bias of the PUF response as well as instability, that is, noise. For this, various methods for evaluating how secure a PUF is have been presented. Among the most popular methods are inter- and intra-Fractional Hamming Distance (FHD) \cite{Maiti2009}, as well as the NIST test suite for random and pseudorandom number generators \cite{Andrew10}. These methods do not require evaluating directly the underlying probability mass function according to which a PUF response is drawn. On the other hand, they do not provide a single measure that quantifies the interplay between noise and bias in terms of the security level; inter-FHD distance is related to bias whereas intra-FHD can be related to noise, and so when using these measures it is not clear whether a PUF with $49\%$ inter-FHD and $10\%$ intra-FHD is more secure than a PUF with $45\%$ inter-FHD and $5\%$ intra-FHD; the NIST test suite is extremely sensitive to bias and does not take into account the effect of noise. Other methods that rely on evaluation of the underlying probability mass function are the min-entropy \cite{FuzzyExt2007}, mutual information \cite{CoverBook} and guesswork \cite{Wang16}. These methods indeed incorporate the effect of noise and bias into a single quantifiable measure.  

\subsection{Contributions}
\begin{itemize}
\item We implement stable PUFs using randomness extracted from the plasma induced oxide breakdown and the voltage stressed oxide breakdown.
\item Test structures violating antenna rules are fabricated with 65nm CMOS bulk technology. Measured results from 99 testchips show that the responses are highly stable across combinations of voltage (0.8V, 1.0V, 1.2V) and temperature variations (25\textdegree C, 100\textdegree C). Compared to a practical SRAM PUF, significant area reduction can be achieved from eliminating ECC implementation for the highly stable responses.
\item We analyze the data from these testchips and show based on various statistical distance measures that pairs of bits with the same antenna ratio as well as bits that are located next to each other are effectively statistically independent. We also propose to use guesswork as a new measure for statistical distance that has operational meaning in terms of security. 
\item We discuss various methods for evaluating the security level of PUFs such as min-entropy, guesswork, inter- and intra-FHD as well as the NIST test suite. In addition, we present the merits of guesswork as a method to evaluate the security level of a PUF. Furthermore, we present the tradeoff between hardware size, bias and guesswork based on our measured results from 99 testchips.

\end{itemize}

\vspace{-0.2in}
\section{Stable PUFs Using Gate Oxide Breakdown}\label{GB}
In this section, we first introduce the gate oxide breakdown and describe two approaches exploiting the gate oxide breakdown as randomness sources of stable PUFs, followed by PUF bit generation and attack resilience analysis.
\vspace{-0.2in}
\subsection{Gate Oxide Breakdown}

Gate oxide breakdown is detrimental to metal-oxide-semiconductor (MOS) devices because it can cause significant drifts of transistor parameters. The breakdown can be categorized into two types: soft breakdown and hard breakdown, where both mechanisms introduce significant sudden increase of the leakage current. For soft breakdown, the conducting path from gate to the substrate is formed by the charged traps in the gate oxide. Once there is conduction, new traps begin to accumulate due to thermal damage, which in turn increases the conductance. The positive feedback eventually leads to thermal runway and oxide is physically melt in the breakdown spot. This type of breakdown is called hard breakdown. The gate leakage current of an oxide with both soft and hard breakdown can be 100X larger than the leakage current of an oxide without breakdown.

\subsection{Plasma Induced Gate Oxide Breakdown} 
During silicon wafer fabrication, plasma processes are widely used for etching, photoresist stripping, or ion implantation. In the plasma ambient, metal segments, VIAs, or polysilicon electrodes, which are the antenna segments, can be electrically charged by ions or electrons, and therefore produce the antenna voltage. For the antenna segments connected to the gate inputs, the resulting electrical stress from the antennas can potentially damage the underlying gate oxide and create a conducting path from the gate to the substrate. The phenomenon is called plasma induced gate oxide breakdown, or the antenna effect. 


Though the maximum voltage rise can be modeled, the actual voltage still cannot be predicted because the exact motion and amounts of ions and electrons collected by the antenna segment are random and unpredictable. The higher the gate voltage is, the higher the probability for the gate oxide breakdown to occur, thus causing a device to fail. Also, systematic plasma variation across wafer does not have much impact on the local randomness because the variation is negligible to a die. 

To avoid the antenna effect, design rules of the antenna ratio (AR) as shown in equation (\ref{eq:AR}) must be strictly followed during fabrication. Practical design rules of AR range from 100 to 5000 depending on the process details.

\begin{equation} \label{eq:AR}
  AR = \frac{\text{exposed antenna area}}{\text{gate oxide area}}
\end{equation}

Since both soft breakdown and hard breakdown can induce about 100X or more leakage current than a good oxide, they are both considered as breakdown in our proposed stable PUF construction. Since the process parameters of our testchip fabrication are unknown prior manufacturing, we implemented a variety of antenna ratios to measure breakdown probabilities, which are presented in Section \ref{sec:exp}. While foundries try to avoid antenna effect during manufacturing, we exploit the uncontrollable physical phenomena as another randomness source of a stable PUF.

\vspace{-0.2in}
\subsection{Voltage Stressed Gate Oxide Breakdown}
The purpose of antenna rules is to protect all transistors from having deviated parameters, for example 20\% gate leakage increase at 1.4xVDD \cite{PJLiao13}, which could be harmful for a normal fabrication but still far from causing a real breakdown. Therefore, to introduce a noticeable plasma induced breakdown (100X increase of leakage current) with 50\% probability of a transistor, an AR larger than 1000X antenna rule may be required, which can result in large area overhead.

To avoid using large antenna segments, one way is to apply high voltage stress to the gate of a transistor directly. By voltage stressing the gate terminal of a transistor, oxide breakdown can be introduced with small AR or even without violating the antenna rules. On the other hand, such a PUF construction requires an additional stress step post manufacturing (or during PUF enrollment). Please note the voltage stressed gate oxide breakdown mechanism is different from the Erasable PUF proposed in \cite{ruhrmair2011attack}, where oxide breakdown is introduced to erase targeted bit cells instead of being used as a stable source of randomness.

\vspace{-0.2in}
\subsection{Stable Signal Unit Construction}
\begin{figure}[h]
\begin{centering}
\includegraphics[width=3.5in]{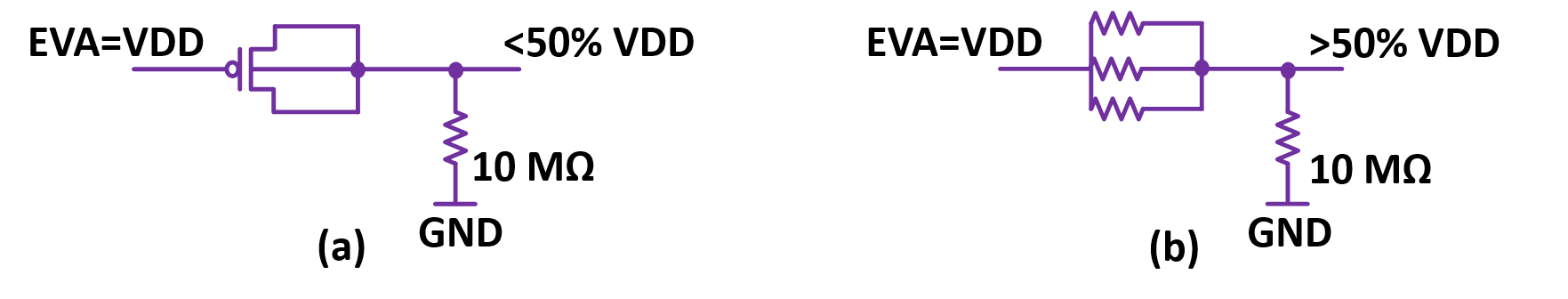}
\par\end{centering}
\caption{Schematic of antenna SSU attached to a precision resistor.}
\label{fig:SSU_unit}
\end{figure}

The permanent gate oxide breakdown mechanism, which can be caused by plasma damage or voltage stressed damage, is used to construct a Stable Signal Unit (SSU) as a source of permanent defectivity. A SSU is a p-MOS transistor designed to violate antenna rules, and its drain, source, and bulk terminals are connected to capture the effect of the gate oxide breakdown at all possible locations. Similar to a gate oxide breakdown model given in \cite{kimreliable}, the SSU is attached in series to a precision resistor as given in Fig. \ref{fig:SSU_unit}, where Fig. \ref{fig:SSU_unit} (a) shows a SSU without oxide breakdown and Fig. \ref{fig:SSU_unit} (b) shows a SSU with oxide breakdown. If no breakdown occurs as depicted in Fig. \ref{fig:SSU_unit} (a), the device is essentially a capacitor or a resistor much larger than the precision resistor, thus the output voltage would be lower than 50\% VDD when the evaluation signal EVA is VDD; if a breakdown happens, as shown in Fig. \ref{fig:SSU_unit} (b), the device can be seen as resistors much smaller than the precision resistor, thus the output voltage would be higher than 50\% VDD when EVA is VDD. The resistance of the precision resistor (10M$\upOmega$) is determined by actual measurements from 99 testchips as described in Section \ref{sec:exp}.

\vspace{-0.2in}
\subsection{Attack Resilience}
 It is worth mentioning that the SSU is more secure than an antifuse cell because an antifuse cell is programmed with hard breakdown only, while the output of the SSU is decided by both soft breakdown and hard breakdown, and a soft breakdown is much harder to detect than a hard breakdown (albeit possible for a very resourceful attacker). For probing attack, the efficiency is limited by the mechanical constraints. For imaging attacks, such as Scanning Electron Microscopy (SEM), Transmission Electron Microscopy (TEM), or Electron Beam Induced Currents (EBIC), it is difficult to efficiently identify a soft breakdown because a soft breakdown because its physical appearance is very similar to a fresh gate oxide without any visible holes. It is also challenging for EBIC to identify a soft breakdown because the limited current of a soft breakdown can induce measurement noises, and the throughput of the electron beam is low. Finally, it is also difficult to observe a soft breakdown from a top-down or cross-section TEM because the image does not effectively tell the depth of the traps. 
 

\section{Testchip Fabrication and Measurement} \label{sec:FM}

\subsection{SSU Implementations}

The proposed SSUs are implemented and fabricated on 99 testchips with commercial 65nm GP 1P9M\_6X1Z1U CMOS bulk technology with 1V nominal voltage. The smallest gate size (0.0072$\mu m^{2}$) of the technology is used for all the SSUs. In our testchips the fabricated SSUs intentionally violate antenna rules by a few hundred times to a few thousand times on different layers. 

On each chip, 29 SSUs are implemented with 17 different ARs, therefore the total number of SSU implementations is 2871 from 99 chips. For each of the SSUs, the cell area and detailed antenna violation report are given in Table \ref{table:ratios}, where a zero indicates that there is no antenna rule violation on such layer. The antenna rule violation reports are provided to the foundry to skip such design rule checks. The M\_T, V\_T, and P\_T structures test the effects of metal, VIA, and polysilicon layers from small AR to large AR, respectively. For each of the M\_T, V\_T, and P\_T, two SSUs with same AR are implemented, therefore 24 bits of responses are obtained from these SSUs on a chip. The remaining five test structures are of various combinations of the violating layers, and one SSU is implemented for each of the five test structures. In summary, on each chip, 29 bits are measured, and 24 bits of them are obtained from the duplicated 12 structures of M\_T, V\_T, and P\_T. 


\begin{table}
\centering
\caption{Cell area, accumulated areas of VIA, metal, polysilicon, and polysilicon perimeter of SSUs fabricated. The numbers are in $\mu m^{2}$ except for the Poly Perimeter ($\mu m$). A zero indicates no antenna rule violation.}
\begin{tabular}{|c|c|c|c|c|c|c|}
\hline
  & Cell  &  VIA  &  Metal  & Poly  & Poly Perimeter\\
\hline
M\_T1 &36 & 0.87 & 1144.57 & 0.00 & 0.00 \\ \hline
M\_T2 &360 & 1.17 & 1468.57 & 0.00 & 0.00 \\ \hline
M\_T3 &1200 & 0.00 & 4398.88 & 0.00 & 0.00 \\ \hline
M\_T4 &4800 & 0.16 & 36781.89 & 0.00 & 0.00 \\ \hline
V\_T1 &2.4 & 0.87 & 1108.57 & 0.00 & 0.00 \\ \hline
V\_T2 &8 & 2.31 & 1108.57 & 0.00 & 0.00 \\ \hline
V\_T3 &90 & 15.27 & 1185.66 & 0.00 & 0.00 \\ \hline
V\_T4 &804 & 144.91 & 1895.05 & 0.00 & 0.00 \\ \hline
P\_T1 &4.8 & 1.26 & 1917.53 & 0.00 & 0.00 \\ \hline
P\_T2 &27 & 1.26 & 1917.53 & 18.17 & 55.59 \\ \hline
P\_T3 &203 & 1.26 & 1917.53 & 180.07 & 128.43 \\ \hline
P\_T4 &1800 & 1.26 & 1917.53 & 1800.07 & 222.46 \\ \hline
Test1 &804 & 1071.86 & 5631.11 & 0.00 & 0.00 \\ \hline
Test2 &4.7 & 1.86 & 0.00 & 0.00 & 0.00 \\ \hline
Test3 &80 & 0.26 & 299.20 & 0.00 & 0.00 \\ \hline
Test4 &60 & 20.84 & 318.78 & 28.07 & 83.81 \\ \hline
Test5 &118 & 54.40 & 617.25 & 56.39 & 164.72 \\ \hline
\end{tabular}
\label{table:ratios}
\end{table}
\subsection{Breakdown Probability Evaluation} \label{sec:exp}
\begin{figure}[htb]
\centering
\includegraphics[width=3.1in]{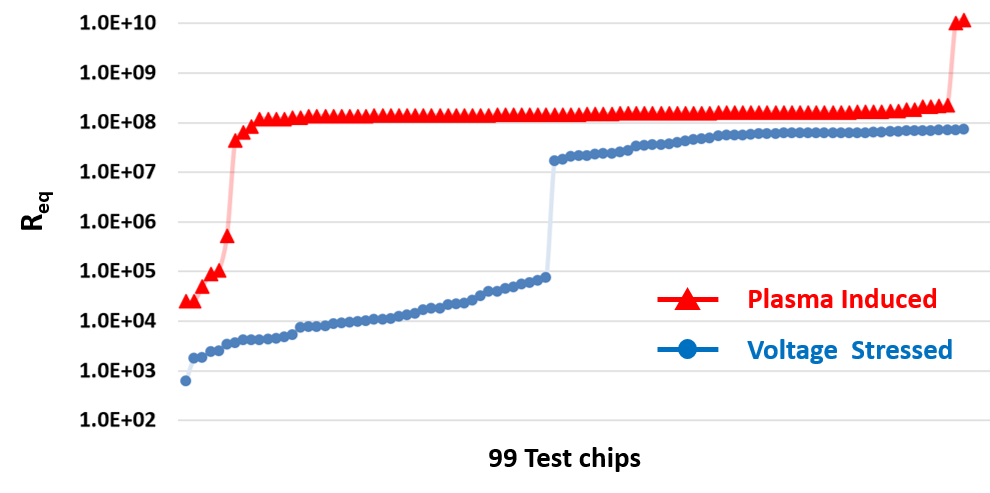}
\caption{The $R_{eq}$ distribution of a SSU implementation (V\_T1) with plasma induced and voltage stressed oxide damage on 99 chips at 25\textdegree C, 1V.}

\label{figure:distribution}
\end{figure}

To determine the gate oxide breakdown of a SSU, we use Agilent 34411A Digital Multimeter to measure the equivalent resistance $R_{eq}$ of each SSU, and from the distribution of $R_{eq}$ we choose a proper precision resistor as shown in Fig. \ref{fig:SSU_unit} to determine whether or not an oxide breakdown has occurred. Fig. \ref{figure:distribution} shows $R_{eq}$ distribution of a SSU implementation (V\_T1) with plasma induced and voltage stressed breakdown on 99 chips in an increasing order at 25\textdegree C, 1V. For both distributions, the $R_{eq}$ of a SSU implementation with oxide breakdown is at least 100X smaller than a SSU without oxide breakdown. After voltage stress, the $R_{eq}$ are in general smaller and much more oxide breakdowns are introduced. The results are similar for all SSUs. The large gap in the figure can be effectively exploited to generate stable digital signals from SSUs. Therefore, we choose, according to the $R_{eq}$ measurements, a 10M$\upOmega$ precision resistor to measure the gate oxide breakdown of each SSU.  

For the plasma induced breakdown, the results of breakdown probabilities of SSU implementations on 99 chips are shown in Table \ref{table:fail_prob}. From the table we see that the breakdown probability of each SSU after plasma induced oxide damage is well below 50\%. This means the responses of SSUs are highly biased, which is undesirable for its low randomness in each response bit. Using larger AR to further increase the breakdown probability may not be a proper approach due to large area overhead.

For the voltage stressed breakdown, we stress 24 SSUs (M\_T, V\_T, and P\_T groups) on each testchip by applying 5.5V to the EVA for 10 seconds. The results of the stress are shown in Table \ref{table:fail_prob}. From the table we can see that breakdown probabilities, which are only slightly correlated with the ARs, are elevated to at least 50\% even for the SSUs with the smallest ARs. These results show that more unbiased responses compared to plasma induced breakdown can be achieved by using small SSUs such as V\_T1. Therefore, a SSU can be implemented with much smaller area, possibly even without violating the antenna rule, than the plasma induced breakdown approach.

\begin{table}
\centering
\caption{Breakdown probability of 17 AR implementations on 99 testchips.}

\begin{tabular}{|c|c|c|}
\hline
 & Plasma Induced & Voltage Stressed\\
\hline
M\_T1 & 0.5\% &  57.6\% \\ \hline
M\_T2 & 0.5\% &  51.5\% \\ \hline
M\_T3 & 2.5\% &  57.1\% \\ \hline
M\_T4 & 2.0\% &  51.0\% \\ \hline

V\_T1 & 0.5\% &  50.0\% \\ \hline
V\_T2 & 6.1\% &  54.0\%  \\ \hline
V\_T3 & 0.0\% &  64.7\%  \\ \hline
V\_T4 & 0.0\% &  58.6\%  \\ \hline

P\_T1 & 1.0\% &  50.5\%\\ \hline
P\_T2 & 2.5\% &  51.5\%\\ \hline
P\_T3 & 1.0\% &  58.6\%\\ \hline
P\_T4 & 1.0\% &  60.0\%\\ \hline

Test1 & 16.2\% & N/A \\ \hline
Test2 & 2.0\% & N/A \\ \hline
Test3 & 5.1\% & N/A \\ \hline
Test4 & 1.0\% & N/A \\ \hline
Test5 & 3.0\% & N/A \\ \hline

\hline

\end{tabular}
\label{table:fail_prob}
\end{table}
\vspace{-0.1in}
\subsection{Stability Evaluation}
To evaluate the stability of the SSUs, we measure all SSU responses from 99 chips at 6 corners: temperatures at 25\textdegree C and 100\textdegree C with $\pm20\%$ voltage variation at 0.8V, 1V, and 1.2V. 
\vspace{-0.1in}
\subsubsection{Plasma Induced Breakdown}
\begin{figure}[htb]
\centering
\includegraphics[width=3.5in]{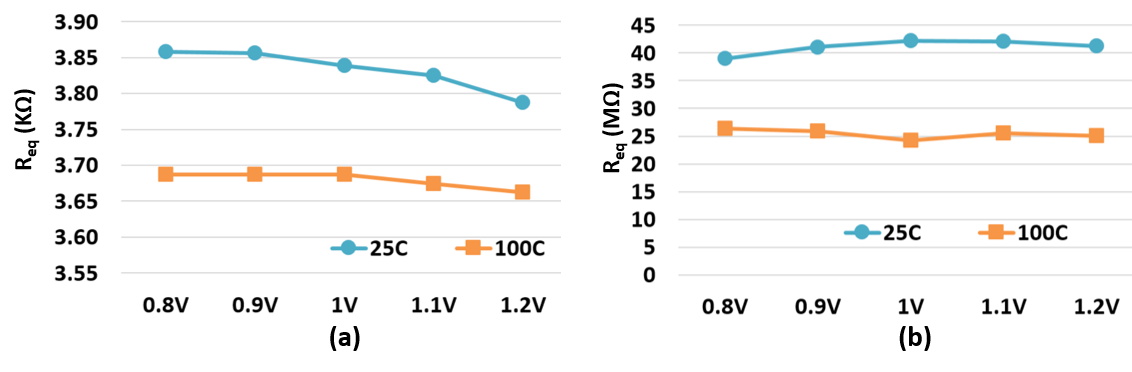}
\caption{Equivalent resistance under extreme voltage and temperature variations. (a) SSU with oxide breakdown. (b) SSU without oxide breakdown.}
\label{figure:Req}
\end{figure}

For the plasma induced breakdown, all SSUs from 99 chips (total 2871 bits generated) are completely stable at all corners during multiple measurements. This can be explained by the fact that the change of $R_{eq}$ at different corners are limited. Fig. \ref{figure:Req} shows the change of $R_{eq}$ of a SSU (Test1) under voltage and temperature variations. In Fig. \ref{figure:Req} (a), the $R_{eq}$ of the SSU with breakdown is only a few K$\upOmega$ and the changes under extreme temperature and voltage variations are limited. On the other hand, Fig. \ref{figure:Req} (b) shows a SSU without oxide breakdown, where the $R_{eq}$ remains at less than 45M$\upOmega$, which is still orders of magnitude larger than the SSU with oxide breakdown. 
\vspace{-0.2in}
\subsubsection{Voltage Stressed Breakdown}
Unlike the plasma induced breakdown, for the voltage stressed breakdown, an extremely small portion of the SSUs are not completely stable. To quantize the results of stability evaluation for the voltage stressed breakdown, each SSU is measured 10 times at each corner and we define the responses measured at 25\textdegree C with 1V, where all responses are consistent, as the reference responses. A SSU is unstable at a corner if at least one of its values from the 10 measurements is different from the reference response. We define bit error rate (BER) the number of unstable bits divided by 2376, which is the total number of SSUs stressed (24 SSUs on each of the 99 chips). Table \ref{table:error_prob} shows the numbers of unstable SSUs and BER at each corner. We found that at several corners, 1 to 3 SSUs out of 2376 SSUs implemented are unstable for the voltage stressed breakdown. Since most responses of unstable SSUs are still consistent with the reference responses, taking the majority vote of multiple measurements can effectively eliminate the erroneous responses.

\begin{table}
\centering
\caption{Bit Error Rates of 2376 SSUs of the voltage stressed breakdown at 6 corners.}

\begin{tabular}{|c|c|c|c|}
\hline
 Corners & 0.8V & 1V & 1.2V \\ \hline
 25\textdegree C & 0.04\% & 0.00\% & 0.12\% \\ \hline
 100\textdegree C & 0.08\% & 0.08\% & 0.08\% \\ \hline

\end{tabular}
\label{table:error_prob}
\end{table}

\subsection{Uniqueness Evaluation}

\begin{figure}[htb]
\centering
\includegraphics[width=2.1in]{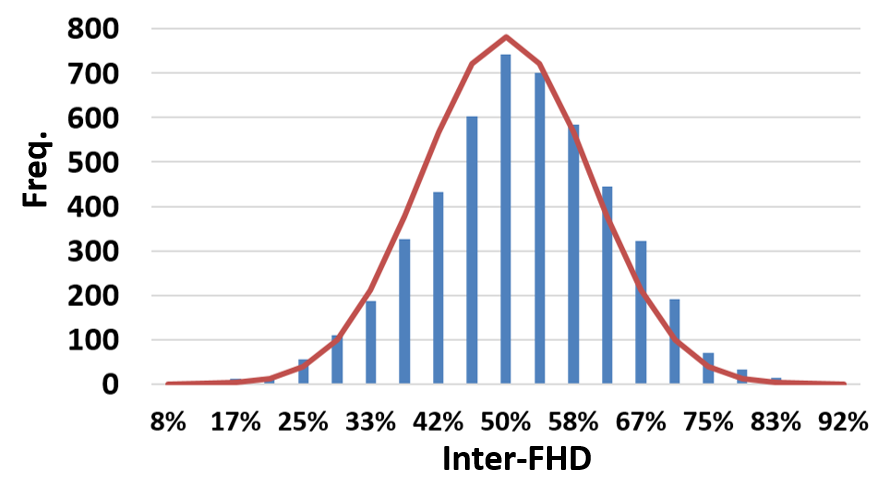}

\caption{Inter-FHD distribution of voltage stressed SSUs on 99 chips overlaid with an ideal Binomial distribution curve with success probability P=0.5.}
\label{fig:inter}
\end{figure}

The inter-Fractional Hamming Distance (FHD) \cite{maiti2013} is calculated as the uniqueness evaluation of SSUs. Consider the 24 voltage stressed SSUs on each chip as a 24-bit weak PUF, the distribution of inter-FHD of 99 chips are presented in Fig. \ref{fig:inter}. The average of inter-FHD is 51.7\% and the standard deviation is 11.4\%, where for an ideal Binomial distribution with success probability P=0.5, the mean is 50\% and the standard deviation is 10.2\%. Please note that the results of uniqueness evaluation are focused on the voltage stressed breakdown SSUs because for the plasma induced breakdown SSUs, the responses are highly biased and post processing would be required to extract randomness, for example using OR gates at the outputs of multiple SSUs to generate an unbiased bit as explained in Section \ref{sec:PPUF}.

\vspace{-0.2in}
\subsection{Statistical Analysis of the PUF Responses}\label{subsec:statisticalanalysisPUF}

We evaluate the statistical dependence between pairs of bits generated by SSUs after voltage stressed oxide breakdown using various statistical distance measures. We consider pairs as we have only $99$ bits per location, and so going beyond the pairwise probability mass function can lead to more noisy and less reliable evaluation. We are interested in the level of independence because the more independent the bits are, the more secure the PUF is. 

Essentially, we use that data to evaluate the pairwise probability mass functions of bits under the following two restrictions: The pairwise probability mass function of bits that have the same antenna ratio; the pairwise probability mass function of bits that are located next to each other. This in turn enables us to evaluate the statistical dependence of element that are more likely to be statistically dependent, that is, statistical dependence due to similar design rules as well statistical dependence between PUFs that are close together. 

We calculate the distance between the evaluated probability mass function (i.e., $P_{X,Y}\pa{x,y}$) and an independent one with the same marginal probability mass functions (i.e., $P_{X}\pa{x}\cdot P_{Y}\pa{y}$) by assigning them to various statistical distance measures. This enables us to demonstrate the level of independence between pairs of bits. The results are presented in Table \ref{table:StatisticalDistance} for the following statistical distance measures: The \textbf{Kullback-Leibler (KL) divergence} \cite{CoverBook} which is defined as
\beq{eq:TheKLDIvergenceDef}{
D\pa{P_{X}||Q_{X}}=\sum_{x\in X}P_{X}\pa{x}\log_{2}\pa{P_{X}\pa{x}/Q_{X}\pa{x}} 
}
and \textbf{total variation distance (TVD)} \cite{kennedy1989}
\beq{}{
\delta\pa{P_{X},Q_{X}}=\frac{1}{2}\sum_{x\in X} \lvert P_{X}\pa{x}-Q_{X}\pa{x}\rvert.
}
 
Table \ref{table:StatisticalDistance} shows that the average statistical distance between $P_{X,Y}\pa{x,y}$ and $P_{X}\pa{x}\cdot P_{Y}\pa{y}$ is very small across measures; note that the maximum value of both of these measures is $1$ for binary random variables, and that these statistical distances are equal to zero when the probability mass functions are identical. Hence, these results indicate that this PUF response is very close to being statistically independent. 

Note that there are many other statistical distance measures that can be used for this purpose and here we provide only a sample of two of the most popular ones. The KL divergence measures the distance between two probability functions in terms of the increase in the average length of codewords when compressing a source which is optimal under $P_{X}$ according to $Q_{X}$, whereas the total variation distance is equal to normalized distance between two probability functions in terms of the $\mathbb{L}_{1}$ norm. However, the operational meaning of these statistical distances as well as many others cannot be directly related to security; in Subsection \ref{subsec:GuessworkAsAStatisticalDistanceMeasure} we propose guesswork as a new statistical distance measure that has an operational meaning from security perspective. 

\begin{table}
\centering
\caption{Statistical distances based on the collected data. In each entry the left side represents the statistical distance of bits that are located next to each other, whereas the right side represents the distance of bits that have the same antenna ratio.}
\begin{tabular}{|c|c|c|c|}
\hline
Statistical Distance & Max & Min & Mean \\
\hline
KL  & $0.11/0.057$ & $0.0002/0.0001$ & $0.022/0.015$\\
\hline
TVD & $0.19/0.13$ & $0.009/0.007$ & $0.07/0.05$ \\
\hline

\end{tabular}\label{table:StatisticalDistance}
\end{table}
\vspace{-0.1in}
\section{Gate Oxide Breakdown PUF Implementations}

\subsection{Plasma Induced Breakdown PUF} \label{sec:PPUF}

To reduce the bias in this structure, we propose to use OR gates at the output of SSUs as a more area-efficient approach than using even larger antenna segments, which shows limited impact on increasing the breakdown probability. Fig. \ref{fig:implementation} (a) shows an exemplary implementation of plasma induced breakdown PUF. The on-chip 10M$\upOmega$ precision resistor is shared between two SSUs, where only one of EVA$_{1}$ and EVA$_{2}$ will be asserted. Please note that a precision resistor can be shared by more than two SSUs to reduce the effective area required per bit, but only one of the SSUs is asserted at a time. The outputs of buffer gates are determined by the breakdown of the SSU.

Take Test3 as an example. When 11 Test3 SSUs are ORed together, the probability of generating a zero is $(1-5.1\%)^{11} = 56\%$, and the area is 880$\mu m^{2}$, which is still more area-efficient than a practical SRAM PUF implementation where (511,19,119)-BCH is suggested to correct 15\% error probability at different corners \cite{Guajardo07}. For such SRAM PUF to generate 19 information bits, the estimated BCH implementation is 12000 XOR gates \cite{Xinmiao2015} or an area of 54000$\mu m^{2}$ for the 65nm technology we used. To generate the same number of 19 bits of response with Test3, the estimated area is about 16720$\mu m^{2}$. The comparison shows that the \textit{SRAM PUF is more than 3X of size of the plasma induced breakdown PUF}. In addition, the ECC execution latency is eliminated for the plasma induced breakdown PUF. 

\subsection{Voltage Stressed Breakdown PUF} \label{sec:SPUF}

\begin{figure}[htb]
\centering
\includegraphics[width=3.4in]{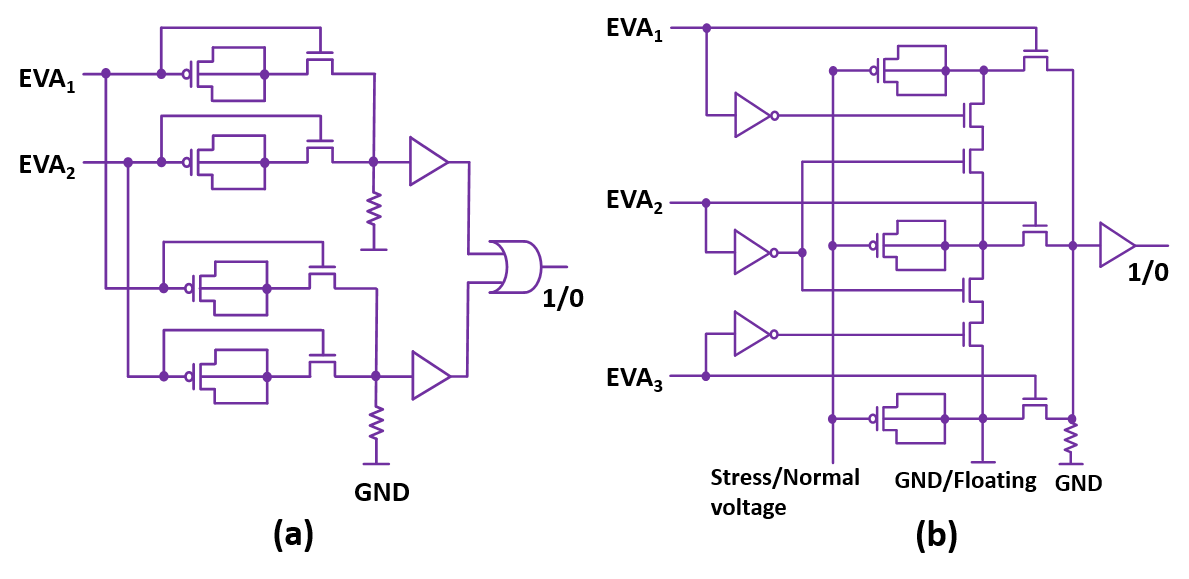}

\caption{(a) Plasma induced breakdown PUF implementation. (b) Voltage stressed breakdown PUF implementation.}
\label{fig:implementation}
\end{figure}

The probability of voltage stressed breakdown is much higher than the plasma induced breakdown, therefore no OR gates are needed to reduce the response bias, but a stress path for each SSU is required. Fig. \ref{fig:implementation} (b) shows an exemplary implementation of voltage stressed breakdown PUF. A precision resistor is shared by 3 SSUs. Before response generation, the PUF is stressed through the stress path and outputs of SSUs are connected to GND with all EVA signals set to zero. Once SSUs are stressed, a normal voltage is applied to the stress path and one of the EVAs is asserted at a time for evaluation. To generate a bit, approximately 1 inverter and 4 transistors are needed, which translates to an area of only 4$\mu m^{2}$ for 65nm technology. The PUF can be stressed on chip, for example with a charge pump with an area overhead of 12200$\mu m^{2}$. Therefore, to generate 19 bits of response, the total area is approximately 12276$\mu m^{2}$, which is about 30\% smaller than the plasma induced breakdown PUF. As the number of bits increases, the area reduction becomes more evident since the charge pump is shared among multiple bits. The PUF can also be stressed from outside of the chip to save even more area, but an antifuse cell may be needed at the stress path. To stress the PUF, the antifuse cell has to be permanently programmed to closed state. Therefore, if the antifuse cell is already in closed state before stress, it means that the PUF has been contaminated and should be discarded. Please note that if the PUF is stressed from outside of the chip, an attacker may destroy the PUF or introduce more breakdowns by further stressing the PUF, but the PUF is not programmable or clonable because the breakdown of each transistor cannot be controlled.


\vspace{-0.2in}
\section{Guesswork for Evaluating Security}

A PUF is expected to provide a certain security level; PUFs are implemented in hardware and so it is desired to minimize the hardware size required to achieve this security level. Therefore, accurate tools for evaluating the security level as a function of the hardware size are needed. Guesswork has been suggested as a measure for the security level of PUFs \cite{Wang16} by connecting PUF security to the framework of password security. In this section we present guesswork along with other measures for the security level of PUFs, and discuss differences between those measures. 

In many scenarios it is reasonable to assume that an attacker can have multiple guesses through which he can try to find a PUF response or alternatively learn its structure. Guesswork can be used in order to evaluate the security level under various types of attacks of this sort such as key stretching \cite{WagnerKeyStreatching}, the guesswork of strong PUFs when bias is presented (e.g., when a model building attack enables an attacker to better predict the response of the next challenges), and the average number of guesses for various probabilities of attack failure.

In this context it is important to note that in terms of the attack model, it is assumed that the attacker can generate responses to challenges based on the statistical profile of the PUF, but does not have access to the device itself (i.e., it is not doing model building). Essentially it means that in this section we focus on attacks against weak PUFs rather than machine learning attacks against strong PUFs. Finally, we also propose guesswork as a new measure for statistical distance that quantifies from security perspective how close random variables are to being statistically independent.
\vspace{-0.2in}
\subsection{Some Background}
The inherent random signature
in the hardware determines how hard it is to guess the response of a weak PUF. The number of guesses required to correctly find the response is termed guesswork \cite{Arikan_Ineq_Guessing}, \cite{Messy1994} and denoted by $G\pa{X}$, where $X$ is a random variable whose response is guessed. $G\pa{X}$ is also a random variable where the probability of having $G\pa{x}$ guesses is $P_{X}\pa{x}$. The $\rho$th moment of guesswork is 
\beq{eq:BasicGuesswork}{
E\pa{G\pa{X}^{\rho}}=\sum_{x}G\pa{x}^{\rho}\cdot P_{X}\pa{x}\quad \rho> 0.
}
and the $\rho$th moment of the conditional guesswork is defined to be 
\beq{eq:ConditionalGuesswork}{
E\pa{G\pa{X|Y}^{\rho}}=\sum_{y}E\pa{G\pa{X|Y=y}^{\rho}}\cdot P_{Y}\pa{y},
}
where $G\pa{X|Y=y}$ is the guesswork when $Y=y$. Furthermore,
it has been shown \cite{Arikan_Ineq_Guessing} that a dictionary attack, that is, guessing values in descending order in terms of the probability mass function (i.e., the values of $P_{X}\pa{x}$) is optimal in the sense that it minimizes any moment of guesswork as well as maximizes the probability of guessing the right response within a certain number of guesses.

Arikan \cite{Arikan_Ineq_Guessing} presented bounds for the $\rho$th moment of the optimal guesswork, $G^{\ast}\pa{X|Y}$, based on which he showed
that when $\udl{X}$ and $\udl{Y}$ are i.i.d., the exponential growth rate of the optimal guesswork is
\beq{eq:asymptiticconditionalguess}{
\lim_{m\to\infty}\frac{1}{m}\log_{2}\pa{E\pa{G^{\ast}\pa{X|Y}^{\rho}}}=\rho\cdot H_{\frac{1}{1+\rho}}\pa{X|Y}
}
where $m$ is the size of $\udl{X}$ and $\udl{Y}$, $P_{X,Y}\pa{x,y}$ is the probability that $x_{i}=x$, $y_{i}=y$, where $x_{i}$, $y_{i}$ are the $i$th elements in $\udl{X}$ and $\udl{Y}$ respectively, and  
\beq{}{
H_{\frac{1}{1+\rho}}\pa{X|Y}=\frac{1}{\rho}\log_{2}\pa{\sum_{y}\pa{\sum_{x}P\pa{x,y}^{\frac{1}{1+\rho}}}^{1+\rho}}} is Renyi's conditional entropy with parameter $\frac{1}{1+\rho}$ \cite{Arikan_Ineq_Guessing}. Note that the average guesswork is achieved when $
\rho=1/2$ in which case the growth rate is equal to $H_{1/2}\pa{X}=2\cdot\log_{2}\pa{\sum_{x}P_{X}\pa{x}^{1/2}}>H\pa{X}=-\sum_{x}P_{X}\pa{x}
\cdot\log_{2}\pa{P_{X}\pa{x}}$ with equality only when $P_{X}$ is the uniform probability mass function, that is, the rate at which the average guesswork increases is larger than the Shannon entropy, which corresponds to guessing over the typical set. 

Note that although equation \eqref{eq:asymptiticconditionalguess} is asymptotic in $m$, that is (i.e., it considers the growth rate of guesswork) the guesswork converges very fast to the exponential term, that is, it converges very fast to $2^{\rho\cdot H_{1/(1+\rho)}\pa{P_{X,Y}\pa{x,y}}\cdot m}$). It is also important to note that just like guesswork, also min-entropy \cite{FuzzyExt2007} and mutual information \cite{CoverBook} represent exponential growth rates that have different operational meanings in terms of security.
\vspace{-0.2in}
\subsection{A Review of Contemporary Security Measures for PUFs}\label{subsec:SurveyonContemporaryMethods}

We divide the methods of evaluating the security level into two groups. Methods that do not require a direct evaluation of the probability mass function, and ones that do require this kind of evaluation.

The main two methods for evaluating the security level without estimating the probability mass function are as follows.
\begin{itemize}
\item \textbf{Inter- and Intra-FHD:} Calculating the inter- and intra-FHD is a very popular method of evaluating the security level \cite{Maiti2009,maiti2013,Roel2010}. The closer the average inter-FHD is to $50\%$ the more unique a PUF is considered to be (i.e., the PUF response is more balanced), whereas the smaller the average intra-FHD distance is, the more reliable the PUF response is, that is, it is more stable. 
\item \textbf{NIST statistical test suite for random and pseudo-random number generators:} NIST offers a statistical test suite \cite{Andrew10} that enables to determine whether or not a string of bits is random enough according to various criteria. This approach can be taken for example when considering PUF based pseudo-random number generators \cite{DevadasPUFPR2004}. 
\end{itemize}

The other two methods for evaluating the security level that rely on evaluating the underlying probability mass function are the following.
\begin{itemize}
    \item \textbf{Mutual Information:} The mutual information between two random variables $X$ and $Y$ (i.e., $I\pa{X;Y}$) quantifies the amount of information that $Y$ reveals on $X$ and vice versa. For example, when $X$ and $Y$ are independent $I\pa{X;Y}=0$, whereas when $X=Y$ then $I\pa{X;Y}=H\pa{X}=H\pa{Y}$. Mutual information has been proposed as a measure for the security level of PUFs \cite{Willems2006,Tuyls2005}. 
    \item \textbf{Min-entropy:} In the context of guessing a secret, the min-entropy \cite{FuzzyExt2007} represents the maximum probability of guessing a secret in a single guess. Since the PUF response is a secret which is chosen at random, min-entropy has also been proposed as a measure for the security level \cite{Katzenbeisser2012}. In the context of machine learning attacks min-entropy has been related to how quickly a strong PUF can be broken \cite{UpperBoundsMinEntropyMLAttacks}. Furthermore, it provides a guarantee for the size of a uniformly distributed key \cite{Dodis2004} that can be extracted.    

\end{itemize}
\vspace{-0.2in}
\subsection{The Merits of Guesswork}\label{sec:TheAdvantageofGuesswork}

In this subsection we explain why guesswork can also be considered as a good measure for the security level of PUFs by going through each of the criteria presented in Subsection \ref{subsec:SurveyonContemporaryMethods} and explaining what the differences between what they evaluate and what guesswork does.

\textbf{Guesswork and inter- and intra-FHD:} The two elements that affect the security level of a PUF are bias and noise. Inter-FHD provides an evaluation of how biased a PUF is and intra-FHD evaluates the noise level, but yet they do not provide a single quantifiable measure that enables one to accurately evaluate the interplay between the two. This in turn may lead to an \emph{inaccurate} evaluation of the security level that might result in either designing a PUF of excessive \emph{size} or a PUF that \emph{does not meet} the required security level.

On the other hand, Guesswork incorporates the effect of bias and noise into a single measure that enables us to evaluate how much they affect the security level. This interplay is presented for the $\rho$th moment of guesswork in  \cite{YonaDiggaviWCTIFSArxiv} by the following equation
\bal{eq:BoundsAsymptoticCondGesswork}{\resizebox{0.92\hsize}{!}{$
\lim_{m\to\infty}\frac{1}{m}\log_{2}\pa{E\pa{G^{\ast}\pa{X,N}^{\rho}}}=\max\pa{\rho H_{\frac{1}{1+\rho}}\pa{X}-\rho\cdot H\pa{N},0},$
}}
where $G^{\ast}\pa{\udl{X},N}$ is the optimal guesswork when $\udl{X}$ is drawn \iid Bernoulli$\pa{p}$, where $p\le 1/2$, and the samples of the PUF encounter an additive noise which is drawn \iid  Bernoulli$\pa{N}$, whose entropy is $H\pa{N}$.
This is proven in \cite{YonaDiggaviWCTIFSArxiv}.

Therefore, when a PUF designer considers the following two PUFs: A PUF with inter-FHD $50\%$ and intra-FHD $10\%$; another PUF with inter-FHD $45\%$ and intra-FHD $5\%$, he can not determine which one is more secure based on inter- and intra-FHD. On the other hand, by assigning the bias and level of noise to equation \eqref{eq:BoundsAsymptoticCondGesswork} he can see that when the noise level is $5\%$ and bias is $35\%$  (which leads to inter-FHD $45\%$ when the bits are i.i.d.) the average guesswork is equal to $2^{128\cdot 0.68}\approx 2^{87}$, whereas when the noise is $10\%$ and the bias is $50\%$ the average guesswork equals $2^{128\cdot 0.53}\approx 2^{68}$. Therefore, guesswork enables a PUF designer to determine which one is more secure and by how much; in this example the PUF with inter-FHD $35\%$ and intra-FHD $5\%$ is $2^{19}$ times more secure than the other one.  

\textbf{NIST statistical test suite:} The main disadvantage of NIST statistical test suite is that it does not designed to take into consideration the effect of noise on the security level. Therefore, a PUF with inter-FHD $50\%$ and intra-FHD $0\%$ (i.e., a stable PUF), as well as a PUF with inter-FHD $50\%$ and intra-FHD $50\%$ (i.e., a maximally noisy PUF) will both pass the NIST test suite.

Moreover, it determines whether or not a string is sufficiently random compared to strings that are drawn \iid Bernoulli$\pa{1/2}$. Therefore, when there is any inherent bias (e.g. the bits are drawn \iid Bernoulli$\pa{1/2-\eps}$) the PUF does not pass these tests. However, the effect of small bias is extremely small in terms of guesswork. This can be seen by assigning a small bias to equation \eqref{eq:BoundsAsymptoticCondGesswork}; for example, when the bits of a stable PUF are drawn \iid Bernoulli$\pa{0.45}$ the average guesswork is equal to $2^{128\cdot 0.9964}\approx 2^{127}$. 

\textbf{Mutual information and min-entropy:} 
In many cases, an attacker can have multiple guesses in which he tries to find the PUF response or learn its structure. Guesswork provides a framework through which a PUF designer can evaluate the security level under such attacks.

Similarly to guesswork, mutual information and min-entropy both incorporate noise and bias into a single expression. \textbf{In fact, in the context of finding the correct response to a challenge, mutual information and min-entropy are both special cases of guesswork}; hence, from this perspective, guesswork can be viewed as a generalization of these two methods. 

\emph{Min-entropy} is the exponent of the maximum probability that the number of guesses is equal to $1$ which is also \textbf{the probability that the optimal guesswork is equal to $1$}, that is, in the \iid case the min-entropy is 
\beq{eq:DefinitionofMinEntropy}{\resizebox{0.92\hsize}{!}{$
H_{min}\pa{X}=-\log_{2}\pa{\max\pa{P_{X}}}=-\frac{1}{m}\log_{2}\pa{Pr\pa{G\pa{X}=1}}.$
}}

In addition, it is shown in \cite{YonaDiggaviWCTIFSArxiv} that min-entropy also captures the average guesswork for strong PUFs under model building attacks.

In \cite{YonaDiggaviWCTIFSArxiv} it is also shown that the \emph{mutual information} between the initial PUF observation and a noisy one, is the average guesswork when guessing across the typical set \cite{CoverBook}, that is, the most probable set. In this case the probability that a PUF response is outside this set is very close but not equal to $1$, that is, the average number of guesses is approximately $2^{128\cdot\pa{I\pa{X;Y}+\eps}}$ when the probability of attack failure is $2^{-128\cdot\eps}$, where $\eps\ll 1$. Therefore, when it comes to guessing, mutual information is the same as guessing with a very small probability of attack failure; this is again a special case of guesswork. The bounds in equation \eqref{eq:BoundsAsymptoticCondGesswork} are achieved when the attacker actually stops when he successfully guesses the secret, even if he has to go though all possibilities.

\subsection{The Impact of Noise on the security level} \label{sec:level_security}
In this subsection we evaluate the impact of noisy responses in terms of the average guesswork of a fixed length response; we consider guesswork for the same reasons provided in Subsection \ref{sec:TheAdvantageofGuesswork}. We compare the voltage stressed breakdown PUF presented in Section \ref{sec:SPUF}, whose noise level is 0.12\%, to show that it performs better than some weak PUFs that have been reported in the literature \cite{Guajardo07} in terms of the average number of guesses required to break them. For a 128-bit response generated by a PUF, we define \textbf{the number of effective bits} as $\pa{H_{1/2}\pa{X}-H\pa{N}}\times 128$, that is, the effective number of bits according to which the exponent of the average guesswork increases when there is bias and noise as defined in equation \eqref{eq:BoundsAsymptoticCondGesswork}. 

\begin{figure}[htb]
\centering
\includegraphics[width=2.6in]{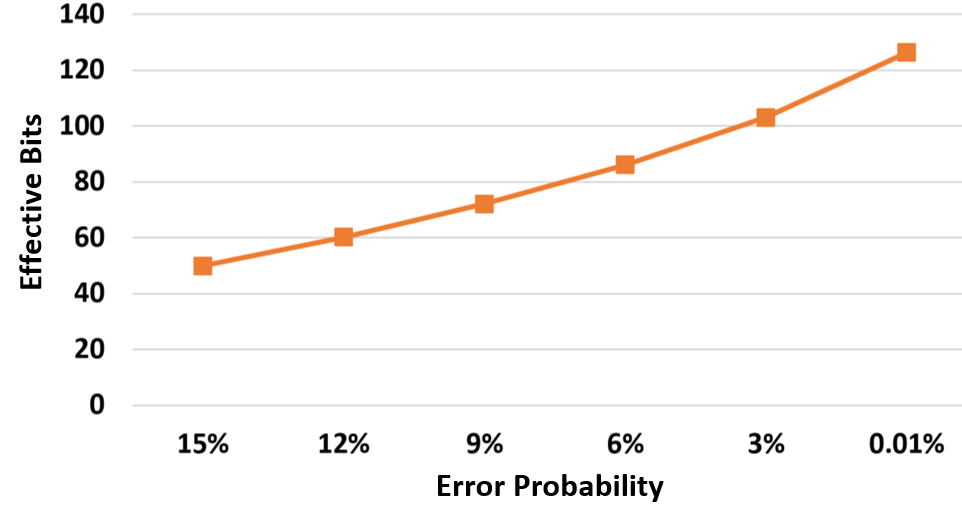}

\caption{Number of effective bits of a 128-bit responses with different error probabilities.}
\label{fig:security_level}
\end{figure}

For the voltage stressed breakdown PUF, $\pa{H_{1/2}\pa{X}-H\pa{N}}$ is 0.9866 as calculated from equation \eqref{eq:BoundsAsymptoticCondGesswork} for its 0.12\% noise level at a worst corner, therefore its number of effective bits is $0.9866\times 128=126.3$, which gives an average guesswork of approximately $2^{126}$. For a weak PUF with 15\% error probability, the number of effective bits is only $0.3902\times 128 = 49.9$ with an average guesswork of approximately $2^{50}$. This means that our new proposed PUFs present a significant improvement at this level in terms of the average guesswork. Figure \ref{fig:security_level} shows the number of effective bits for various noise levels. We can see that the number of effective bits drops dramatically as the noise level increases, which reflects how severe the noise can affect the security level in terms of the average guesswork.

\vspace{-0.2in}
\subsection{Guesswork as a Statistical Distance Measure}\label{subsec:GuessworkAsAStatisticalDistanceMeasure}

In contrast with the previous subsections, which used guesswork to evaluate the security level, in this subsection we propose \textbf{guesswork as a new statistical measure} that enables one to measure the statistical distance between random variables. The advantage of using guesswork as a statistical distance measure over other measures such as those presented in Subsection \ref{subsec:statisticalanalysisPUF} lies in the fact that the number associated to this distance has operational meaning in terms of security.

The new statistical distance is defined as follows:
\bal{}{
G\pa{P_{X,Y},P_{X}\cdot P_{Y}} = \frac{\log_{2}E\pa{G\pa{X}^{\rho}}-\log_{2}E\pa{G\pa{X|Y}}^{\rho}}{\log_{2}E\pa{G\pa{X}}^{\rho}}
\nonumber\\=
\pa{H_{1/\pa{1+\rho}}\pa{X}-H_{1/\pa{1+\rho}}\pa{X|Y}}/H_{1/\pa{1+\rho}}\pa{X}.}
When $G\pa{P_{X,Y},P_{X}\cdot P_{Y}}= \alpha$ it means for large $m$ that 
\beq{}{
E\pa{G\pa{X|Y}^{\rho}}=\pa{E\pa{G\pa{X}^{\rho}}}^{1-\alpha} .
}
Hence, when $X$ and $Y$ are independent $\alpha=0$ and so $E\pa{G\pa{X}^{\rho}}=E\pa{G\pa{X|Y}^{\rho}}$, whereas $X=Y$ leads to $\alpha=1$ as well as $G\pa{X|Y}=1$ . We present $G\pa{P_{X,Y},P_{X}\cdot P_{Y}}$ for our evaluations of $P_{X,Y}$ and the marginals $P_{X}$, $P_{Y}$ when $\rho=1$ (i.e., the average guesswork) in Table \ref{table:StatisticalDistanceGW}. These results show that when $X$ is conditioned on $Y$ there is a loss of about $1\%$ on average in terms of the exponent of $E\pa{G\pa{X}}$, that is, when $E\pa{G\pa{X}}=2^{m}$, $E\pa{G\pa{X|Y}}=2^{0.99\cdot m}$. Therefore, the statistical statistical dependence between $X$ and $Y$ is very weak in terms of guesswork.

\begin{table}
\centering
\caption{In each entry the left side represents the statistical distance of bits that are located next to each other, whereas the right side represents the distance of bits that have the same antenna ratio.}
\begin{tabular}{|c|c|c|c|}
\hline
Statistical Distance & Max & Min & Mean \\
\hline
GW  & $0.06/0.029$ & $0.0001/0.00009$ & $0.011/0.008$\\
\hline
\end{tabular}\label{table:StatisticalDistanceGW}
\end{table}

\vspace{-0.2in}
\section{Conclusion}
In this paper we implement and analyze highly stable PUFs exploiting uncontrollable plasma induced and voltage stressed gate oxide damage. The proposed SSUs are fabricated and measured from 99 testchips. Measurement results show that the SSUs are highly stable, therefore significant area reduction can be achieved by eliminating ECC implementation. Furthermore, we show that the responses are unbiased and unique, and we analyze the data of our testchips using various statistical distance measures to show that these bits are independent. Finally, we present the merits of guesswork as a measure for evaluating the security level of PUFs.

\end{document}